\documentclass[bm,aps,showpacs,amsfonts,amssymb]{revtex4}  

\usepackage{graphicx}
\usepackage{natbib}
\begin{document}

{~}
\title{
Uniqueness theorems for Kaluza-Klein black holes\\
 in five-dimensional minimal supergravity
}
\vspace{2cm}
\author{Shinya Tomizawa\footnote{E-mail:tomizawa@post.kek.jp}}
\vspace{2cm}
\affiliation{
${}^{1}$Cosmophysics Group, Institute of Particle and Nuclear Studies, \\
KEK, Tsukuba, Ibaraki, 305-0801, Japan 
${}^{2}$Department of Mathematics and Physics,Graduate School of Science,Osaka City University, 3-3-138 Sugimoto, Sumiyoshi, Osaka 558-8585, Japan
}
\begin{abstract} 
We show uniqueness theorems for  Kaluza-Klein black holes in 
the bosonic sector of five-dimensional minimal supergravity. 
More precisely, under the assumptions of the existence of two commuting axial 
isometries and a non-degenerate connected event
horizon of the cross section topology $S^3$, or lens space, we prove that
 a stationary charged rotating Kaluza-Klein black hole in five-dimensional minimal supergravity 
is uniquely characterized by its mass, two independent angular momenta, electric charge,
magnetic flux and nut charge, provided that there does not exist any nuts in the domain of outer communication. 
We also show that under the assumptions of the same symmetry, same asymptotics and 
the horizon cross section of $S^1\times S^2$, a black ring within the same theory---if exists---is uniquely determined by its dipole charge and rod structure besides the charges and magnetic flux.
\end{abstract}

\preprint{KEK-TH 1377}
\pacs{04.50.+h  04.70.Bw}
\date{\today}
\maketitle

\section{Introduction}\label{sec:intro}

Higher dimensional black holes have played an important role in 
understanding basic properties of fundamental theories, such as 
string theory. A number of interesting solutions of higher dimensional 
black holes have been discovered recently~\cite{Myers:1986un,Emparan:2001wn,Pom,diring,saturn,Izumi,bi,MishimaIguchi,Gauntlett0,
BMPV,Elvang,EEMR,Elvang3,CY96,CLP,CCLP,EEF,Y1,Y2,Y3,Y4}, revealing much 
richer structure of their solution space than that of four-dimensional 
black holes, 
and we are naturally led to address the question of how to classify them. 
There have already appeared several papers that generalize the black
hole uniqueness theorems~\cite{uniqueness,Israel,Israel2,Bunting,Bunting2,Carter,Robinson,Robinson2,Mazur,Bunting3} to higher dimensions~\cite{shiromizu,T-schwarzschild,MI,Hollands,MTY,HY,R,R2,R3,R4,R5,R6,R7,TYI,TYI2},
 upon some additional assumptions concerning the horizon topology,  
symmetry properties, asymptotic structures, etc.
In particular, five-dimensional generalizations of the uniqueness 
theorems have been shown, in various theories,
for stationary, axisymmetric (with two rotational symmetries) 
black holes being non-compact, as a simple higher dimensional generalizations 
of the well-known four-dimensional setup.    
However, since our real, observable world is macroscopically four-dimensional,
extra-dimensions have to be compactified in realistic, classical 
spacetime models~\footnote{ 
The assumption of asymptotic global flatness becomes relevant 
in the context of a certain type of braneworld models,  
in which size of higher dimensional black holes can be much 
smaller than the size of extra-dimensions.
}.   
Therefore it is of great interest to consider higher dimensional 
Kaluza-Klein black holes, which look like four-dimensional, at least 
at large distances. 
Classifying such Kaluza-Klein solutions may also help us to get some insights
into the major open problem of how to compactify and stabilize
extra-dimensions in string theory. The purpose of this paper is to 
address such a classification problem, showing a uniqueness theorem for 
stationary Kaluza-Klein black hole solutions of five-dimensional minimal 
supergravity. 

\medskip

Perhaps the simplest example of Kaluza-Klein black holes is 
a black-string, a direct product of a four-dimensional vacuum black hole
and a circle. 
A more non-trivial class of Kaluza-Klein black holes is given by
{\sl Squashed} Kaluza-Klein black holes, found recently  
by Ishihara-Matsuno~\cite{IM}, applying the squashing technique to
five-dimensional black holes. The idea is that for, e.g., the simplest  
static vacuum case, one first views the $S^3$ section 
(or horizon manifold) of a five-dimensional Schwarzschild-type black hole 
spacetime as a fibre bundle of $S^1$ over $S^2$, and then considers 
a deformation that changes the ratio of the radii of the fibre $S^1$ 
and base $S^2$, so that the resultant spacetime looks, 
at large distances, like a twisted $S^1$ over a four-dimensional asymptotically
flat spacetime, hence a Kaluza-Klein spacetime, while it looks like 
a five-dimensional black hole near the event horizon.
The basic structure of squashed Kaluza-Klein black holes can in fact
be seen in the much earlier works of refs.~\cite{DM,Rasheed}, whose 
solutions asymptote to a twisted $S^1$-bundle over a four-dimensional 
spacetime as studied in~\cite{GW}. 
A number of further generalizations of squashed Kaluza-Klein black
holes has been made lately~\cite{Bena2,Gaiotto,IKMT,NIMT,TIMN,MINT,TI,TYM,GS}. The recent accumulation of this new 
type of Kaluza-Klein black hole solutions also motivates us to address the 
classification problem of Kaluza-Klein black holes.

\medskip
All known exact Kaluza-Klein black hole solutions in five-dimensions 
admit the isometry group that describes the stationarity 
and two `axial' symmetries, one along $S^2$ base space and the other along 
$S^1$ fibre, or simply $T^2$. 
These symmetries are mutually commuting, hypersurface orthogonal,  
and form the isometry group ${\rm R}\times U(1) \times U(1)$. 
In this paper, we consider Kaluza-Klein black holes that
possess this symmetry property and that are purely bosonic as 
solutions to the minimal supergravity. 
The topology of horizon cross-sections can be either $S^3$, 
$S^1 \times S^2$, or lens space $L(p,q)$\footnote{ 
Topology censorship in Kaluza-Klein spacetimes~\cite{Chru} tells that
topology of $T^3$ is not allowed under the assumption of null energy condition. }.
More precisely we shall show the following:  

\medskip
\medskip
\medskip 
\noindent
{\bf Theorem.} 
{\em 
Consider the bosonic sector in five-dimensional minimal supergravity, i.e., in five-dimensional Einstein-Maxwell-Chern-Simons theory 
with a certain special value of the Chern-Simons coupling constant [given by eq.~(\ref{action}) below], a stationary charged rotating 
black hole with finite temperature that is regular on and outside 
the event horizon and asymptotically Kaluza-Klein spacetime [the precise definition is given below].
Assume that the black hole spacetime admits, besides the stationary Killing 
vector field, two mutually commuting axial Killing vector fields so that 
the isometry group is ${\Bbb R}\times U(1)\times U(1)$ and that the topology of the horizon spatial cross-sections is either $S^3$, or $L(n;1)$, or
$S^1 \times S^1$.
Then 
(i) if there exists no nut in the domain of outer communication, the black hole spacetime with $S^3$, or $L(n;1)$ horizon cross-section is uniquely characterized
by its mass, and two independent angular momenta, electric charge, nut charge and magnetic flux, and
(ii) if the topology of the black hole exterior region is ${\Bbb R} \times \{ {\Bbb R}^4 \setminus B^3\times S^1 \}$, the black hole spacetime with $S^1\times S^2$ horizon cross-section is uniquely characterized
by its dipole charge and rod structure in addition to their charges and magnetic flux.
} 

\medskip
\medskip
\medskip

\medskip
It is known that the bosonic sector of minimal supergravity that possess 
the above symmetry group can be reduced to a non-linear sigma 
model~\cite{MO,MO2}, which is much more complicated than the well-known 
four-dimensional electrovacuum case. Nevertheless, one can derive formulas 
similar to those used in the four-dimensional uniqueness proof, such as 
the coset matrix representation of the equations of motion, 
the divergence (Mazur) identity, etc.,~\cite{BCCGSW}, as we will discuss below. 
Apart from the difference in the sigma-model, another main difference 
from the four-dimensional case can be seen in the boundary value analysis, 
in particular, along the symmetry axis and the horizon. 
This is because we have a larger variety of the horizon topology 
in five-dimensions. We can specify the horizon topology 
in terms of the `rod-structure' (or interval structure)~\cite{Harmark}.

\medskip 
The new part of the job that is particular for our asymptotic Kaluza-Klein 
case is in the boundary value analysis at infinity. 
We need to consider fall-off conditions of the sigma-model 
fields at infinity more carefully than the globally flat case.  
When inspecting the asymptotic fall-off behavior of the perturbations, 
we find that the perturbations of the metric and the gauge field 
decouple each other at least in the leading order. Then, imposing 
boundary conditions for the asymptotically Kaluza-Klein spacetime, 
we will identify the parameters which are used to characterize the solutions. 
For example, from the leading order of the fall-off behavior of 
the gravitational sector, we find $N$ which describes how much 
the $S^1$-fibre (i.e., the compactified $5$th-dimension) 
is twisted with respect to the $S^2$ base space, 
and $Q$ which may be viewed as the `angular' momentum along $5$th-dimension, 
in addition to the usual angular momentum $J$ along $U(1)$ of the base space. 
From the Maxwell part, we have, besides the electric charge $q$,
the magnetic flux $c_\phi$ over the base space at infinity, and furthermore
the dipole charge $q_m$ 
if the topology of the horizon cross section is $S^1\times S^2$
(See the next section for their precise definitions.).

\medskip 
The main interest of this paper is in the context of minimal supergravity, 
and we therefore restrict our attention to the Einstein-Maxwell-Chern-Simons (EMCS) 
theory with a certain value of the coupling constant. 
However, one can expect that a similar uniqueness theorem may also hold 
in other similar theories. For example, restricting attention to some 
integrable sector of the five-dimensional pure Einstein-Maxwell 
theory, a uniqueness theorem similar to the above has 
recently been shown~\cite{Y10}, based on the classification of 
Kaluza-Klein black holes in arbitrary, $D$-dimensional, vacuum Einstein 
gravity~\cite{HY08} with $D-2$ Killing symmetries. 
The sigma-model for the integrable sector of five-dimensional pure 
Einstein-Maxwell system appears quite different, but the basic strategy 
for the proof is essentially the same as the one we will consider below. 
However, we should note that the integrable sector considered in~\cite{Y10} corresponds to a highly restricted class of the solutions 
in which the electric part of the Maxwell field and, at least, one of the 
two angular momentum are required to vanish. 
For this reason, the boundary value analysis in the asymptotic region 
(i.e., at large distances) in~\cite{Y10} 
seems rather simple and straightforward. 
Furthermore, for the solutions dealt with 
in~\cite{Y10}, some of the parameters in our above theorem turn 
out to be identically zero. 

\medskip 
In this paper we consider more (perhaps the most) general 
class of Kaluza-Klein black hole solutions with a single horizon of the minimal supergravity 
that possess the above isometry group, so that the solutions can admit 
two independent arbitrary `angular' momenta (one of which may be 
called the momentum along the compactified dimension) 
and non-vanishing electric component of the Maxwell field. 
We find that for such a general solution, for example, 
the parameter $c_\phi$ appears, in contrast to the case of asymptotically flat cases~\cite{TYI,TYI2} 
for which $c_\phi$ vanishes.    
Note also that for some known exact solutions~\cite{Gauntlett0,GH,Herdeiro,TIMN,MINT,TI}, the parameter $c_\phi$ seems to be related to so-called G\"odel parameter, whose
 square is proportional to the energy density of magnetic field. 
As well known~\cite{Gauntlett0,GH,Herdeiro},  the five-dimensional G\"odel type universe is filled with the pressureless magnetic field 
and due to the rapid rotation of the magnetic field, the spacetime admits closed timelike curves in far regions, but for Kaluza-Klein black holes~\cite{TIMN,MINT,TI} it exhibits no causal pathology outside
the event horizon, thanks to the appropriate compactification.
As far as we know, 
the most general solution with all independent parameters has not yet been found.
 Hence, we would like to show that the such a solution is characterized by their parameters and therefore must be unique {\it if it exists}.

\medskip
In the next section, we will briefly describe our strategy for 
the proof and write down some necessary formulas, such as 
the equations of motion, the definitions of relevant 
sigma-model fields. 
In Section~\ref{sec:infinity}, by solving straightforwardly EMCS equations near infinity, we derive the Kaluza-Klein asymptotics
---the asymptotic behaviors of the metric and gauge potential of Maxwell-Chern-Simons field----in the Weyl-Papapetrou coordinate system.
In Section~\ref{sec:bh} we perform the boundary value analysis for black holes with a spherical horizon topology and 
complete our proof. In Section \ref{sec:br} we consider the boundary value analysis for black rings with $S^1\times S^2$ horizon cross section
 and show the uniqueness theorem. In Section \ref{sec:bl} we also discuss the boundary value analysis for black lenses. In 
Section \ref{sec:summary} we summarize our results.

\section{Einstein-Maxwell-Chern-Simons system with symmetries}  

\subsection{The basic strategy for the proof and mathematical formulas}
\label{sec:metric}

First we briefly describe the basic strategy for our uniqueness proof, 
which roughly proceeds as follows. 
(i) We first reduce the (bosonic sector of)
five-dimensional minimal supergravity theory with three commuting 
independent Killing symmetries to a non-linear sigma model, that is, 
set of equations for eight scalar fields $\Phi^A$ on two-dimensional
orbits space $\Sigma$, with the target space isometry $G$.
With the aid of $G$, the action of the sigma model can be described
in terms of a symmetric, unimodular matrix, $M$, on the coset 
space $G/H$ where $H$ is an isotropy subgroup of $G$. Thus, the solutions
of the original system can be expressed by the matrix $M$.
Furthermore, the matrix $M$ formally defines 
a conserved current, $J$, for the solution. 
(ii) Next, we introduce the {\em deviation matrix}, $\Psi$, which is 
essentially the difference between two coset matrices, 
say $M_{[0]}$ and $M_{[1]}$, so that when two solutions coincide with 
each other, the deviation matrix vanishes, and vice versa. 
What we wish to show is that $\Psi$ vanishes over the entire $\Sigma$ 
when two solutions satisfy the same boundary conditions that specify relevant 
physical parameters characterizing the black hole solution of interest. 
For this purpose, we construct a global identity, called the 
{\em Mazur identity}, (the integral version of) which equates an integration 
along the boundary $\partial \Sigma$ of a derivative of the trace of $\Psi$ 
to an integration over the whole base space $\Sigma$ of the trace of 
`square' of the deviation, ${\cal M}$, of the two conserved currents, 
$J_{[0]}$ and $J_{[1]}$. The latter is therefore non-negative. 
(iii) Then, we perform boundary value analysis of the matrix $\Psi$.   
We identify boundary conditions for $M$ that define physical parameters 
characterizing black hole solutions and that guarantee the regularity 
of the solutions. 
Then we examine the behavior of $\Psi$ near $\partial \Sigma$. 
For higher dimensional case, this is the point where the topology and symmetry 
properties, translated into the language of the rod-structure, come to play 
a role as additional parameters to specify solutions.
Also this is the place where we have to take into consideration
the nature of asymptotic structure of the spacetime. 
When the integral along the boundary $\partial \Sigma$, say the left-side 
of the Mazur identity, vanishes under our boundary conditions, it then 
follows from the right-side of the identity, i.e., the non-negative 
integration over $\Sigma$, that ${\cal M}$ has to vanish, hence 
the two currents, $J_{[0]}$ and $J_{[1]}$, must coincide with each other 
over $\Sigma$, 
implying that the deviation matrix $\Psi$ must be constant over $\Sigma$. 
Then, if $\Psi$ is shown to be zero on some part of the boundary 
$\partial \Sigma$, it follows that $\Psi$ must be identically zero 
over the entire $\Sigma$, thus proving the two solutions, 
$M_{[0]}$ and $M_{[1]}$, must be identical. 

\medskip 
In our present case, the first two steps (i)-(ii) completely parallel 
those in Paper \cite{TYI}, and Step (iii) is the new result of this paper. 
In the following we provide some of the formulas for Steps (i) 
and (ii), such as the definitions of the relevant sigma-model fields, 
in order to establish our notation. The reader can also find them 
in Paper~\cite{TYI}.  
Some relevant formulas, such as the coset matrix representation 
of the sigma-model field, are also summarized in the appendix.

\subsection{Einstein-Maxwell-Chern-Simons system with symmetries 
and the reduction to $\sigma$-model} 
\label{sec:metric}

We start with the five-dimensional minimal supergravity action 
\begin{eqnarray}
S=\frac{1}{16\pi}
  \left[ 
        \int d^5x\sqrt{-g}\left(R-\frac{1}{4}F^2\right) 
       -\frac{1}{3\sqrt{3}} \int F\wedge F\wedge A 
  \right] \,, 
\label{action} 
\end{eqnarray} 
where we set a Newton constant to be unity and $F=dA$. 
Varying this action (\ref{action}), we derive the Einstein equation 
\begin{eqnarray}
 R_{\mu \nu } -\frac{1}{2} R g_{\mu \nu } 
 = \frac{1}{2} \left( F_{\mu \lambda } F_\nu^{ ~ \lambda } 
  - \frac{1}{4} g_{\mu \nu } F_{\rho \sigma } F^{\rho \sigma } \right) \,, 
 \label{Eineq}
\end{eqnarray}
and the Maxwell equation 
\begin{eqnarray}
 d*F+\frac{1}{\sqrt{3}}F\wedge F=0 \,,  
\label{Maxeq}
\end{eqnarray}
which have the extra term coming from the Chern-Simons term of (\ref{action}). 
We are concerned with asymptotically Kaluza-Klein, stationary, charged rotating 
black hole solutions of this theory. We additionally impose two independent 
axial symmetries, so that the total isometry group is 
${\Bbb R}\times U(1) \times U(1)$ with ${\Bbb R}$ being stationary 
symmetry, generated by mutually commuting three Killing vector fields 
$\xi_t = \partial /\partial t$ and $\xi_a= (\xi_\phi, \xi_w)
=(\partial/\partial \phi, \partial/\partial w)$\footnote{
This assumption concerning {\it two} independent axial symmetries is 
not fully justified, as the rigidity theorem~\cite{HIW,MI08,HI} 
guarantees the existence of only a single axial symmetry for 
stationary black holes.  
}. 
Using the Einstein equations and the Maxwell equations, we can show that 
the generators $\xi_t ,\: \xi_a$ of the isometry group satisfy 
type of integrability conditions discussed in Ref.~\cite{Harmark,weyl}. 
As a result, we obtain the coordinate system, $\{t,\phi,w,\rho,z\}$, 
in which the metric takes the {\em Weyl-Papapetrou form} 
\begin{eqnarray}
ds^2&=&\lambda_{\phi\phi}(d\phi+a^\phi{}_tdt)^2
     + \lambda_{ww}(dw+a^w{}_tdt)^2 
\nonumber\\ 
   &&+2\lambda_{\phi w}(d\phi+a^\phi{}_tdt)(dw+a^w{}_tdt)
     +|\tau|^{-1}[e^{2\sigma}(d\rho^2+dz^2)-\rho^2 dt^2] \,, 
\end{eqnarray}
and the gauge potential is written, 
\begin{eqnarray} 
 A = \sqrt{3}\psi_a dx^a + A_t dt \,, 
\label{pote:gauge}
\end{eqnarray}
where the coordinates $x^a=(\phi,w)$ denote the Killing parameters, 
and thus all functions $\lambda_{ab}$, 
$\tau:=-{\rm det}(\lambda_{ab})$, $a^a$, $\sigma$, and 
$(\psi_a,A_t)$ are independent of $t$ and $x^a$, 
and where the potentials $\psi_a$ are related to Maxwell field 
by eq.~(8) of Paper \cite{TYI} 
[see also Appendix A of Paper~\cite{TYI} for the gauge choice employed 
in eq.~(\ref{pote:gauge})]. 
Note that the coordinates $(\rho,z)$ that span a two-dimensional 
{\em base space}, $\Sigma=\{(\rho,z)|\rho\ge 0,\ -\infty<z<\infty \}$, 
are globally well-defined, harmonic, and mutually conjugate on $\Sigma$. 
See e.g., \cite{Chru08}. 
Furthermore, by using the Maxwell's equation and Einstein's equations, 
we introduce the magnetic potential $\mu$ and twist potentials $\omega_a$ by   
\begin{eqnarray} 
d\mu&=&\frac{1}{\sqrt{3}}*(\xi_\phi\wedge \xi_w\wedge F) 
    - \epsilon^{ab}\psi_ad\psi_b \,, 
\label{eq:magnetic} 
\\
 d\omega_a&=&*(\xi_\phi\wedge \xi_w \wedge d\xi_a)+\psi_a(3d\mu+\epsilon^{bc}\psi_bd\psi_c)\,, 
\label{eq:twistpotential} 
\end{eqnarray} 
where $\epsilon^{\phi w}=-\epsilon^{w\phi }=1$.
Then, the nonlinear sigma-model reduced from the theory (\ref{action}) with 
the symmetry assumptions consists of the target space with the isometry 
$G=G_{2(2)}$ and the eight scalar fields 
$\Phi^A=(\lambda_{ab},\omega_a,\psi_a, \mu)$ on the base space $\Sigma$. 
All the other fields such as $\sigma,a^a$, etc can be determined by $\Phi^A$ 
through the equations of motion.

\medskip 
It turns out that the sigma model fields, $\Phi^A$, can be expressed 
by a $7 \times 7$ symmetric unimodular coset $G_{2(2)}/SO(4)$ matrix $M$. 
[see eq.~(34) of Paper~\cite{TYI}], as shown by~\cite{MO,MO2,BCCGSW}. 
We will provide the detail description of the coset matrix in Appendix~A.
Then we define the deviation matrix, $\Psi$, for two solutions, 
$M_{[0]}$ and $M_{[1]}$, as in eq.~(42) of Paper \cite{TYI}, 
and derive the Mazur identity, 
\begin{eqnarray}
 \int_{\partial \Sigma}\rho \partial_p {\rm tr} \Psi dS^p 
  = \int_{\Sigma} {\rm tr}({\cal M}^{T}\! \cdot \!{\cal M})\rho d\rho dz \,, 
 \label{eq:mazurid}
\end{eqnarray} 
where {\it dot} denotes the inner product on $\Sigma$. As briefly mentioned 
above, $\cal M$, in the right-side essentially describes the difference 
between two matrix currents $J_{[0]},\:J_{[1]}$, given by eq.~(47) 
of Paper~\cite{TYI}, of which detail is irrelevant to discussion below. 
Our task is to show that the left-side of eq.~(\ref{eq:mazurid}) vanishes on 
the boundary, $\partial \Sigma$, and then show $\Psi$ itself 
vanishes on some part of the boundary.

\medskip 
Now we note that the right-hand side of the identity, (\ref{eq:mazurid}), 
is non-negative. 
Therefore, if we impose the boundary conditions at $\partial\Sigma$,
under which the left-hand side of Eq.(\ref{eq:id}) vanishes, then 
we must have $\stackrel{\odot}J{}^i=0$. In that case, 
it follows from eq.~(\ref{eq:deriv}) that $\Psi$ must be a constant matrix 
over the region $\Sigma$. 
Therefore, in particular, if $\Psi$ is shown to be zero on some part of 
the boundary $\partial \Sigma$, it immediately follows that $\Psi$ must 
be identically zero over the base space $\Sigma$, implying that 
the two solutions $M_{[0]}$ and $M_{[1]}$ must coincide with each other. 
This is indeed the case as we will analyses in the next section.

\section{Kaluza-Klein asymptotics in five dimensions} \label{sec:infinity}

Before estimating the boundary integrals in the left-hand side of the Mazur 
identity, eq.~(\ref{eq:mazurid}), we must derive the asymptotic form of the gauge potential and metric at infinity for asymptotically Kaluza-Klein spacetimes 
including all known exact solutions in $D=5$ minimal supergravity. 
Here, by the {\it asymptotically Kaluza-Klein spacetime}, we mean that the five-dimensional spacetime metric at large distances behaves as
\begin{eqnarray}
ds^2\simeq -dt^2+dx^2+dy^2+dz^2+dw^2
\end{eqnarray}
where the 5-th coordinate $w$ has the periodicity $\Delta w=2\pi L$. Hence, we can see that at infinity, the spacetime behaves as a four dimensional flat spacetime with a circle. Now in order to study the asymptotics of such a spacetime, it is more convenient to use 
the radial coordinate $r$ and the angular coordinate $\theta$ defined by
\begin{eqnarray}
r&=&\sqrt{x^2+y^2+z^2},\\
\theta&=&\arccos\left(\frac{z}{r}\right).
\end{eqnarray}  
Note that the coordinates, $(\rho,z)$, in the Weyl-Papapetrou coordinate system are related to the above defined coordinates, $(r,\theta)$, by
\begin{eqnarray}
\rho&=&r\sin\theta,\\
z&=&r\cos\theta.
\end{eqnarray}

\subsection{Gauge potential}
First, we determine the behavior of the gauge field, $A\simeq A^{(0)}(\theta)+A^{(1)}(\theta)/r+{\cal O}(r^{-2})$, near infinity. From eq. (\ref{Maxeq}), the gauge potential, $A$, is subject to the Maxwell-Chern-Simons equation,
\begin{eqnarray}
\frac{1}{\sqrt{-g}}\partial_\nu \left(\sqrt{-g} F^{\mu\nu}\right)+\frac{1}{4\sqrt{3}}\epsilon^{\mu\nu\rho\sigma\lambda}F_{\nu\rho}F_{\sigma\lambda}=0.\label{eq:max}
\end{eqnarray}
From the $t$-component of eq.(\ref{eq:max}), we can derive the equation to determine the leading order of $A_{t}$,
\begin{eqnarray}
\partial_\theta^2A_t^{(0)}+\cot\theta \partial_\theta A_t^{(0)}=0.
\end{eqnarray}
Solving the above equation, we obtain 
\begin{eqnarray}
A^{(0)}_t=c_t+d_t\log \left|\tan\frac{\theta}{2}\right|,
\end{eqnarray}
where $c_t$ and $d_t$ are integration constants. The regularity of the field strength $F=dA$ requires $d_t=0$. 
Note that by using the gauge transformation, {\it i.e.}, the gauge freedom in adding a constant, 
we can also set the value of the other constant to be $c_t=0$.  
After all, without loss of generality, we may put
\begin{eqnarray}
A^{(0)}_t=0.\label{eq:A0t}
\end{eqnarray}
From the leading order of the $w$-component in eq.(\ref{eq:max}), we derive the equation to the leading order of $A_w$,
\begin{eqnarray}
\partial_\theta^2A_w^{(0)}+\cot\theta \partial_\theta A_w^{(0)}=0.
\end{eqnarray}
Similarly, we get
\begin{eqnarray}
A^{(0)}_w= c_w+ d_w\log \left|\tan\frac{\theta}{2}\right|,
\end{eqnarray}
where $\bar c_w$ and $\bar d_w$ are constants.
From the same discussion, we can set the values of these two integration constants to be $\bar c_w=\bar d_w=0$ and therefore obtain
\begin{eqnarray}
A_w^{(0)}=0.\label{eq:A0w}
\end{eqnarray}
 The remaining $\phi$-component is written as
\begin{eqnarray}
\partial_\theta^2 A_\phi^{(0)}-\cot\theta \partial_\theta A^{(0)}_\phi=0.
\end{eqnarray} 
The solution is written in terms of integration constants $c_\phi$ and $d_\phi$
\begin{eqnarray}
A_\phi^{(0)}=c_\phi\cos\theta+d_\phi.
\end{eqnarray}
Using the degree of the gauge freedom, we may choose $d_\phi=0$ and hence obtain
\begin{eqnarray}
A_\phi^{(0)}=c_\phi\cos\theta.\label{eq:A0phi}
\end{eqnarray}
Substituting eqs.(\ref{eq:A0t}), (\ref{eq:A0w}) and (\ref{eq:A0phi}) into eq.(\ref{eq:max}), we can derive the equations which determine 
the next order of the gauge fields $A^{(1)}$. 
It turns out that the equations for $A^{(1)}_t$ and $A_w^{(1)}$ take the exactly same forms as for $A^{(0)}_t$ and $A_w^{(0)}$.
Therefore, the next orders of $A_t$ and $A_w$ turn out to be, respectively
\begin{eqnarray}
A_t^{(1)}=q,\quad A_w^{(1)}=c_w,
\end{eqnarray}
where $q$ and $c_w$ are constants. Note that by using the gauge transformation, we cannot set them to be $0$.
To summarize, near infinity, the gauge field behaves as
\begin{eqnarray}
A\simeq \frac{q}{r}\left(1+{\cal O}(r^{-1})\right)dt+c_\phi \cos\theta \left(1+{\cal O}(r^{-1})\right)d\phi+\frac{c_w}{r}\left(1+{\cal O}(r^{-1})\right)dw.\label{eq:gauge}
\end{eqnarray}

\subsection{Metric}
Next we would like to determine the next order of the metric, $g_{ij}^{(1)}$, near infinity, 
where $g_{ij}^{(1)}$ is defined by
\begin{eqnarray}
g_{ij}(r,\theta)= g_{ij}^{(0)}(\theta)\sum_{k=1}^\infty\left(1+\frac{g_{ij}^{(k)}(\theta)}{r^k}\right).
\end{eqnarray}
From the $(tt)$-component of eq.(\ref{Eineq}), we can derive the equation to determine the next order of $g_{tt}$,
\begin{eqnarray}
\partial_\theta^2g_{tt}^{(1)}+\cot\theta \partial_\theta g_{tt}^{(1)}=0.
\end{eqnarray}
This can immediately be solved,
\begin{eqnarray}
g_{tt}^{(1)}=c_{tt}+d_{tt} \log\left|\tan\frac{\theta}{2}\right|,
\end{eqnarray}
where $c_{tt}$ and $d_{tt}$ are integration constants.
The regularity of the metric requires $d_{tt}=0$. Hence, we obtain
\begin{eqnarray}
g_{tt}^{(1)}=c_{tt}.
\end{eqnarray}
From the $(ww)$-component and $(tw)$-component of eq.(\ref{Eineq}), we derive the equations, respectively
\begin{eqnarray}
\partial_\theta^2g_{ww}^{(1)}+\cot\theta \partial_\theta g_{ww}^{(1)}=0,
\end{eqnarray}
\begin{eqnarray}
\partial_\theta^2g_{tw}^{(1)}+\cot\theta \partial_\theta g_{tw}^{(1)}=0.
\end{eqnarray}
Similarly, in terms of constants $Q$ and $c_{ww}$, $g_{ww}^{(1)}$ and $g_{tw}^{(1)}$ can be written,
\begin{eqnarray}
g_{ww}^{(1)}=c_{ww},
\end{eqnarray}
\begin{eqnarray}
g_{tw}^{(1)}=Q,
\end{eqnarray}
respectively.
From the $(\phi w)$-component of eq.(\ref{Eineq}), we derive the equation,
\begin{eqnarray}
\partial_\theta^2g_{\phi w}^{(0)}-\cot\theta \partial_\theta g_{\phi w}^{(0)}=0.
\end{eqnarray}
Solving this, we can obtain   
\begin{eqnarray}
g_{\phi w}^{(0)}=c_{\phi w}+N \cos\theta,
\end{eqnarray}
where $c_{\phi w}$ and $N$ are constants. 
It turns out here that by performing the coordinate transformation, $w\to w-c_{\phi w}\phi$, the constant, $c_{\phi w}$, can be set to be $0$. 
Therefore, $g_{\phi w}^{(0)}$ can be written as
\begin{eqnarray}
g_{\phi w}^{(0)}=N \cos\theta.
\end{eqnarray}
From the $(\phi\phi)$-component of eq.(\ref{Eineq}) and the above results, the equation,
\begin{eqnarray}
\partial_\theta^2g_{\phi\phi}^{(1)}+\cot\theta \partial_\theta g_{\phi\phi}^{(1)}=0.
\end{eqnarray}
can be derived.
The regularity of the metric requires that the solution must take the form of
\begin{eqnarray}
g_{\phi\phi}^{(1)}=c_{\phi\phi},
\end{eqnarray}
where $c_{\phi\phi}$ is an integration constant.
From the $(t\phi)$-component of eq.(\ref{Eineq}) , we derive the equation
\begin{eqnarray}
\partial_\theta^2g_{t\phi}^{(1)}-\cot\theta \partial_\theta g_{t\phi}^{(1)}+2g_{t\phi}^{(1)}-2Q N \cos\theta=0.
\end{eqnarray}
Integrating this equation, we obtain the solution,
\begin{eqnarray}
g_{t\phi}^{(1)}=J \sin^2\theta+d_{t\phi}\left(2\cos\theta+\sin^2\theta \log\frac{1+\cos\theta}{1-\cos\theta}\right)+QN\cos\theta,
\end{eqnarray}
in terms of constants $J$ and $d_{t\phi}$. Similarly, 
the regularity requires $d_{t\phi}=0$.

\medskip
Here, recall that in the canonical coordinate system, the three-dimensional metric 
$g=(g_{ij})\ (i,j=t,\phi,w)$ is subject to the constraint 
\begin{eqnarray}
{\rm det}(g)=-\rho^2 \,. 
\end{eqnarray}
Therefore, using the constraint
and the formula,  
\begin{eqnarray}
{\rm det}(g+\delta g) 
 &=&{\rm det}[g(1+g^{-1}\delta g)] 
\nonumber \\ 
 &=&-\rho^2\left(1+{\rm tr}(g^{-1}\delta g)+{\rm det}(g^{-1}\delta g)\right) 
\nonumber \\
 &\simeq&-\rho^2\left(1+{\rm tr}(g^{-1}\delta g)\right) \,, 
\end{eqnarray}
we can see in the next order that the metric has to satisfy the constraint 
\begin{eqnarray}
\sum_{i=t,\phi,w}g^{(1)}_{ii}=0 \,,
\end{eqnarray}
which is the same constraint as in the asymptotically flat case~\cite{TYI}. 
We note that though in the Weyl-Papapetrou coordinate system, the asymptotic form of the metric is not diagonal, 
the off-diagonal component does not affect this constraint in the order of ${\cal O}(r^{-1})$.

\medskip
Thus, to summarize, in the Weyl-Papapetrou coordinate system, the metric near infinity, $r=\sqrt{\rho^2+z^2}\to\infty$, behaves as 
\begin{eqnarray}
ds^2&\simeq& \left(-1+\frac{m}{r}+{\cal O}(r^{-2})\right)dt^2+r^2\sin^2\theta\left(1+\frac{m-\eta}{2r}+{\cal O}(r^{-2})\right)d\phi^2+\left(1+\frac{m+\eta}{2r}+{\cal O}(r^{-2})\right)dw^2\nonumber\\
&&+\frac{2(J\sin^2\theta+Q N \cos\theta)}{r}\left(1+{\cal O}(r^{-1})\right)dtd\phi+\frac{2Q}{r}\left(1+{\cal O}(r^{-1})\right)dtdw+2N\cos\theta\left(1+{\cal O}(r^{-1})\right)d\phi dw\nonumber\\
&&+\left(1+{\cal O}(r^{-1})\right)(d\rho^2+dz^2).\label{eq:metric}
\end{eqnarray}
Here $\eta$ is a constant that comes from 
gauge degrees of freedom in the choice of the coordinate $z$, 
i.e., degrees of freedom with respect to shift translation $z\to z+\alpha$.  
(This gauge freedom exists even after the gauge freedom of 
the conjugate coordinate, $\rho$, is fixed at infinity.) 
Since in our proof we choose the coordinate $z$ such that the horizons 
are located at the interval $[-k^2,k^2]$ for two configurations 
$M_{[0]}$ and $M_{[1]}$, we choose the same values of $\eta$ 
for the two solutions.

\subsection{Asymptotic charges and flux}
Now let us see the relation between the asymptotic charges and the integration constants appearing in asymptotic form of the metric and gauge potential. 
We can see from eq. (\ref{eq:metric}) that for $r\to \infty$, the metric behaves as
\begin{eqnarray}
ds^2\simeq -dt^2+dr^2+r^2(d\theta^2+\sin^2\theta d\phi^2)+(dw+N\cos\theta d\phi)^2,
\end{eqnarray}
It is now clear that the metric has the structure of $S^1$ bundle over the four-dimensional Minkowski space-time
and the spatial infinity is $S^1$ fibre bundle over $S^2$ base space. In particular, when $N=L/2$, or $N=(L/2)n\ (|n|:$ natural numbers larger than one), 
the spatial infinity can be regarded as a squashed $S^3$, or squashed lens space $L(n;1)$. 
Also note when $N=0$, the $S^1$ and the Minkowski spacetime are direct product.
The asymptotic charges should be defined as boundary integrals over such the spatial infinity $S_\infty$. 
Since we are concerned with stationary,
axisymmetric spacetimes with Killing symmetries in EMCS theory, the
conserved charges, mass $M$, angular momenta $J_a$ and electric charge $Q_e$ are defined as follows, and are related to the integration constants in asymptotic form of the metric and gauge potential by 
\begin{eqnarray}
M=-\frac{3}{32\pi} 
        \int_{S_\infty}
                 dS^{\mu \nu} \nabla_\mu (\xi_t)_\nu = \frac{3\pi mL}{4}\,
\label{def:Mass} ,
\end{eqnarray}
\begin{eqnarray}
J_\phi=\frac{1}{16\pi}\int_{S_\infty} 
               dS^{\mu \nu} \nabla_\mu (\xi_\phi)_\nu=\frac{\pi JL}{3} \,
\label{def:Ja} ,
\end{eqnarray}
\begin{eqnarray}
J_w=\frac{1}{16\pi}\int_{S_\infty} 
               dS^{\mu \nu} \nabla_\mu (\xi_w)_\nu=\frac{\pi QL}{2} \,
\label{def:Ja},
\end{eqnarray}
\begin{eqnarray}
Q_e=\frac{1}{16\pi}\int_{S_\infty}\left(*F+\frac{1}{\sqrt{3}}A\wedge F\right)=\frac{\pi qL}{2}.
\end{eqnarray}
As seen later, the magnetic flux $Q_m$ is defined by
\begin{eqnarray}
Q_m=\frac{1}{4\pi}\int_{S_\infty^2}F=c_\phi,
\end{eqnarray}
where $S_\infty^2$ denotes the base manifold of $S^2$ at infinity.

\section{Boundary value problems for black holes} \label{sec:bh}

\begin{figure}[htbp]
 \begin{center}
  \includegraphics[width=40mm]{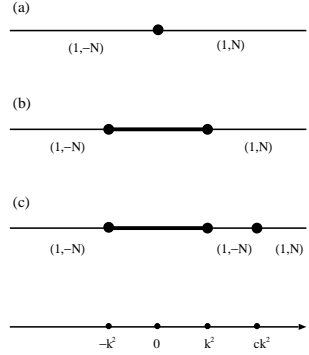}
 \end{center}
 \caption{The rod structures of spacetimes with Kaluza-Klein asymptotics: (a) the Gross-Perry-Sorkin (GPS) monopole, (b) the black hole and (c) the black ring. Here, the solid finite rods correspond to
 the horizons, the assigned vectors on the spacelike rods denote the rod vectors, {\it i.e.}, the pairs of numbers $(1,\pm N)$ means 
that the Killing vectors, $v=(\partial/\partial\phi)\pm N(\partial/\partial w)$, have fixed points there,  more precisely, the metric, $g_{ij}(0,z)$, has an eigenvalue zero for a given $z$. 
See ref.~\cite{Chen-Teo} about the rod structures of well known gravitational instantons --- for example, Euclidean self-dual Taub-NUT space --- with $U(1)\times U(1)$ symmetry and its classification.}
 \label{fig:rod}
\end{figure}

As discussed in~\cite{Hollands,HY08}, under the existence of two commuting axial 
Killing vectors, the cross-section topology of each connected 
component of the event horizon of stationary vacuum black hole solutions 
must be either $S^3$, $S^1\times S^2$ or a lens space. 
First, let us start from the boundary value analysis for black holes with a spherical horizon cross-section and with Kaluza-Klein asymptotics.
In terms of the Weyl-Papapetrou coordinate system 
and the rod-structure \cite{Harmark}, 
the boundary $\partial \Sigma$ of the base space 
$\Sigma=\{(\rho,z)|\ \rho>0,\ -\infty<z<\infty \}$ is described as 
a set of three rods and the infinity (See FIG.\ref{fig:rod}.(b) about the rod diagram):
\begin{enumerate}
\item[(i)]
the outer axis: 
$\partial \Sigma_+=\{(\rho,z)|\rho=0,k^2<z<\infty \}$ with the rod vector 
$v=(0,1,N)$ \,, 
\item[(ii)] the horizon: 
$\partial \Sigma_{\cal H}=\{(\rho,z)|\ \rho=0,-k^2<z<k^2\}$ \,, 
\item[(iii)] 
the outer axis: 
$\partial \Sigma_-=\{(\rho,z)|\rho=0,-\infty<z<-k^2\}$ 
with the rod vector $v=(0,1-N)$ \,, 
\item[(iv)] the infinity:  
$\partial \Sigma_\infty 
= \{(\rho,z)|\sqrt{\rho^2+z^2}\to 
\infty\ {\rm with}\ z/\sqrt{\rho^2+z^2}\ {\rm finite}  \}$ \,, \label{eq:mazur}
\end{enumerate} 
where here and hereafter ${\cal H}$ denotes a spatial cross-section of 
the event horizon. 
As mentioned in the previous section, the sphericity of the spatial infinity requires that the nut charge $N$ must be related to the size of the 5-th dimension $L$ by $N=L/2$.
Furthermore note also that in the above rod structure there is no turning point such as a nut ---the point where two spacelike rods meet with each other--- outside the horizon and therefore
 this means that the topology of the horizon cross section is $S^3$ and the topology of the black hole exterior region is ${\Bbb R} \times \{ {\Bbb R}^4 \setminus {\Bbb B}^4 \}$. 
 We can see this as following.
We here assume the identification $(\phi,w)\to(\phi+2\pi,w+2\pi N)$, $(\phi,w)\to (\phi+2\pi,w-2\pi N)$ and hence the periodicity of $\phi$ and $w$ are $2\pi$ and $4\pi N$, respectively.
 Therefore,  as discussed in ref.~\cite{Chen-Teo}, the pair of Killing vectors, $\partial/\partial\phi_\pm=\partial/\partial\phi\pm N\partial /\partial w$ ($\phi_\pm =(\phi \pm N^{-1}w)/2$), is identified as a pair of
$2\pi$ periodic generators of the $U(1) \times U(1)$ isometry group since the identification $(\phi,w)\to(\phi+2\pi,w+2\pi N)$, $(\phi,w)\to (\phi+2\pi,w-2\pi N)$ in the coordinate $(\phi,w)$ can be regarded as
 the identification of $(\phi_+,\phi_-) \to (\phi_++2\pi,\phi_-)$, $(\phi_+,\phi_-) \to (\phi_+,\phi_-+2\pi)$ in the coordinates $(\phi_+,\phi_-)$. Accordingly, we can see that the deteminant of the two rod vectors $v_\pm:=\partial/\partial\phi_\pm$
 is $|{\rm det}(v_+,v_-)|=1$, which means that the horizon cross section and the spatial infinity are topologically $S^3$.
The boundary integral in the left-hand side of the Mazur 
identity, eq.~(\ref{eq:mazurid}), is decomposed into the integrals over the three 
rods (i)--(iii), and the integral at infinity (iv), as 
\begin{eqnarray}
\int_{\partial \Sigma}\rho\partial_p{\rm tr}\Psi dS^p 
&=& \int_{-\infty}^{-k^2}\rho\frac{\partial {\rm tr}\Psi }{\partial z}dz 
   +\int_{-k^2}^{k^2}\rho\frac{\partial {\rm tr}\Psi }{\partial z}dz 
\nonumber\\
 && +\int_{k^2}^{\infty}\rho\frac{\partial {\rm tr}\Psi }{\partial z}dz 
    +\int_{\partial\Sigma_\infty}\rho\partial_p{\rm tr}\Psi dS^p \,.   
\label{eq:integral} 
\end{eqnarray}

\medskip
(iv) the infinity: 
It immediately follows from eq.(\ref{eq:metric}) that near infinity, the gravitational potentials, $\lambda_{ab}$, behave as
\begin{eqnarray}
&&\lambda_{\phi\phi}\simeq \rho^2\left(1+\frac{m-\eta}{2\sqrt{\rho^2+z^2}}+{\cal O}((\rho^2+z^2)^{-1})\right),\label{eq:l1}\\
&&\lambda_{ww}\simeq 1+\frac{m+\eta}{2\sqrt{\rho^2+z^2}}+{\cal O}((\rho^2+z^2)^{-1}),\label{eq:l2}\\
&&\lambda_{\phi w}\simeq \frac{Nz}{\sqrt{\rho^2+z^2}}+{\cal O}((\rho^2+z^2)^{-\frac{1}{2}})\label{eq:l3}.
\end{eqnarray}
We see directly from eq.(\ref{eq:gauge}) that the electric potentials, $\psi_a$, behave as
\begin{eqnarray}
&&\psi_\phi\simeq \frac{c_\phi}{\sqrt{3}}\frac{z}{\sqrt{\rho^2+z^2}}+{\cal O}((\rho^2+z^2)^{-\frac{1}{2}}),\label{eq:psiphi}\\
&&\psi_w \simeq \frac{c_w}{\sqrt{3}}\frac{1}{\sqrt{\rho^2+z^2}}+{\cal O}((\rho^2+z^2)^{-1}) \label{eq:psiw}.
\end{eqnarray}

From the Kaluza-Klein asymptotics (\ref{eq:metric}), we see that the functions, $\tau$ and $a^a{}_t$, behave as
\begin{eqnarray}
&&\tau =\lambda_{\phi w}^2-\lambda_{\phi\phi}\lambda_{ww} \simeq -\rho^2,\label{eq:tau}\\
&&a^{\phi}{}_t=\frac{\lambda_{\phi w} g_{tw}-\lambda_{ww} g_{t\phi}}{\tau}\simeq -\frac{J}{\sqrt{\rho^2+z^2}},\\
&&a^w{}_t=\frac{\lambda_{\phi w} g_{t\phi}-\lambda_{\phi\phi} g_{t w}}{\tau}\simeq \frac{Q}{\sqrt{\rho^2+z^2}}.\label{eq:aw}
\end{eqnarray}
near infinity. 
From eqs. (\ref{eq:psiphi})-(\ref{eq:aw}), the derivatives of the magnetic potential behaves as
\begin{eqnarray}
&&\mu_{,\rho}=\frac{\tau}{\rho}\left(\frac{A_{t,z}}{\sqrt{3}}-a^a_t \psi_{a,z}\right)-\epsilon^{ab}\psi_a \psi_{b,\rho}\simeq \frac{q\rho z}{\sqrt{3}\sqrt{\rho^2+z^2}^3},\\
&&\mu_{,z}=-\frac{\tau}{\rho}\left(\frac{A_{t,\rho}}{\sqrt{3}}-a^a_t \psi_{a,\rho}\right)-\epsilon^{ab}\psi_a \psi_{b,z}\simeq -\frac{q\rho^2}{\sqrt{3}\sqrt{\rho^2+z^2}^3}\label{eq:muz}.
\end{eqnarray}
Hence, by integrating these, we find that near infinity, the magnetic potential behaves as
\begin{eqnarray}
&&\mu \simeq -\frac{qz}{\sqrt{3}\sqrt{\rho^2+z^2}}+{\cal O}((\rho^2+z^2)^{-1}).\label{eq:mu}
\end{eqnarray}

\medskip
On the other hand, the asymptotic behaviors of the derivatives of the twist potentials can be derived
\begin{eqnarray}
&&\omega_{\phi,\rho}=\frac{\tau}{\rho}\lambda_{\phi b}a^b_{t,z}+\psi_\phi(3\mu_{,\rho}+\epsilon^{bc}\psi_b \psi_{c,\rho})
\simeq \frac{-3J\rho^3z}{\sqrt{\rho^2+z^2}^5}+\frac{(NQ-c_\phi q) \rho z^2}{(\rho^2+z^2)^2},\\
&&\omega_{\phi,z}=-\frac{\tau}{\rho}\lambda_{\phi b}a^b_{t,\rho}+\psi_\phi(3\mu_{,z}+\epsilon^{bc}\psi_b \psi_{c,z})
\simeq \frac{3J\rho^4}{\sqrt{\rho^2+z^2}^5}-\frac{(NQ-c_\phi q) \rho^2 z}{(\rho^2+z^2)^2},\\
&&\omega_{w,\rho}=\frac{\tau}{\rho}\lambda_{wb}a^b_{t,z}+\psi_w(3\mu_{,\rho}+\epsilon^{bc}\psi_b \psi_{c,\rho})
\simeq \frac{Q\rho z}{\sqrt{\rho^2+z^2}^3},\\
&&\omega_{w,z}=-\frac{\tau}{\rho}\lambda_{wb}a^b_{t,\rho}+\psi_w(3\mu_{,z}+\epsilon^{bc}\psi_b \psi_{c,z})
\simeq -\frac{Q\rho^2}{\sqrt{\rho^2+z^2}^3}.
\end{eqnarray}
from its definition (\ref{eq:twistpotential}) and eqs. (\ref{eq:psiphi})-(\ref{eq:aw}).
Integrating these, we obtain
\begin{eqnarray}
&&\omega_\phi \simeq \frac{(c_\phi q-NQ)z^2}{2(\rho^2+z^2)}+J\left(\frac{6z}{\sqrt{\rho^2+z^2}}-\frac{2z^3}{\sqrt{\rho^2+z^2}^3}\right)+{\cal O}((\rho^2+z^2)^{-\frac{1}{2}}),\label{eq:ophi}\\
&&\omega_w\simeq -\frac{Qz}{\sqrt{\rho^2+z^2}}+{\cal O}((\rho^2+z^2)^{-\frac{1}{2}}).\label{eq:opsi}
\end{eqnarray}

\medskip
Then, using eqs.(\ref{eq:l1})-(\ref{eq:l3}), (\ref{eq:psiphi}), (\ref{eq:psiw}), (\ref{eq:mu}), (\ref{eq:ophi}) and (\ref{eq:opsi}), we can see that for the two configurations, $M_{[0]}$ and $M_{[1]}$, with the same constants, $(m,J,Q,N,q,c_{\phi})$, $\rho\ {\rm tr}\Psi$ near infinity behaves as  
\begin{eqnarray}
{\rm tr}\Psi
&\simeq& \frac{6\left(\stackrel{\odot}c_w\right)^2}{\rho^2+z^2}\,. 
\end{eqnarray}
Therefore, by using the coordinates $(r,\theta)$, $\rho\ \partial_p{\rm tr}\Psi dS^p$ at infinity $r=\infty$ turns out to be
\begin{eqnarray}
 \rho\ \partial_p{\rm tr}\Psi dS^p\simeq 6\left(\stackrel{\odot}c_w\right)^2  \left(r\sin\theta \right)\cdot\left( \partial_{\theta}\ r^{-2}\right)\cdot (rd\theta)=0 \,,\label{eq:inf}
\end{eqnarray}
which does not depend on whether $\stackrel{\odot}c_w$ vanishes, or not.
Thus, we can show that for the two solutions with the same values of the constants $(m,J,Q,N,q,c_{\phi})$, the boundary integral at infinity vanishes
\begin{eqnarray}
\int_{\partial\Sigma_\infty}\rho\ \partial_p{\rm tr}\Psi dS^p=0.
\end{eqnarray}

\vspace{1cm}

\medskip
(ii) the horizon: 
$\partial \Sigma_{\cal H}=\{(\rho,z)|\ \rho=0,-k^2<z<k^2\}$. The regularity on the horizon requires that for $\rho \to 0$,   
\begin{eqnarray}
&&\lambda_{ab}\simeq {\cal O}(1),\quad \omega_{a}\simeq {\cal O}(1) \,,\\
&&\psi_{a}\simeq {\cal O}(1), \quad \mu\simeq {\cal O}(1) \,. 
\end{eqnarray}
Therefore, for $\rho \to 0$, $\rho\ {\rm tr}\Psi$ behaves as
\begin{eqnarray}
\rho\ \partial_z{\rm tr}\ \Psi\simeq O(\rho) \,.  
\end{eqnarray}

\medskip
(i), (iv) the outer axises: 
$\partial \Sigma_\pm=\{(\rho,z)|\rho=0,k^2<|z|<\infty \}$. We note that the rod vectors $v=(1,\pm N)$ vanish on the two outer axises. By just the same discussion as in the black lens~\cite{TYI2}, 
the regularity requires that for $\rho \to 0$, the potentials $\lambda_{ab}$ must behaves as
\begin{eqnarray}
&&\lambda_{\phi\phi}\simeq N^2g(z)+{\cal O}(\rho^2), \label{eq:olambda11}\\
&&\lambda_{\phi w}\simeq \mp Ng(z)+{\cal O}(\rho^2), \label{eq:olambda12}\\
&&\lambda_{ww}\simeq g(z)+{\cal O}(\rho^2), \label{eq:olambda22}
\end{eqnarray}
where $g(z)$ is some function of $z$. Note here that in the below boundary value analysis, one need not require $g(z)_{[0]} =g(z)_{[1]}$ for the two solutions with the same boundary condition.
\medskip 
Next, let us consider the boundary conditions for the electric potentials $\psi_a$. 
It follows that for $\rho=0$,
\begin{eqnarray}
0=-i_vF=\sqrt{3}(d\psi_\phi\pm Nd\psi_w) \,. 
\end{eqnarray}
Integrating this, we obtain
\begin{eqnarray}
\psi_\phi\pm N\psi_w=c_0 \,, 
\label{eq:c0}
\end{eqnarray}
where $c_0$ is a constant. Note from eqs.~(\ref{eq:psiphi}) and (\ref{eq:psiw}) that $c_0= \psi_\phi(\rho=0,z=\pm \infty)\pm N\psi_w(\rho=0,z=\pm \infty)=\pm\frac{c_\phi}{\sqrt{3}}$.
Therefore, we can set the electric potentials to behave as 
\begin{eqnarray}
&&\psi_\phi\simeq \pm \frac{c_\phi}{\sqrt{3}}\mp N h(z)+{\cal O}(\rho^2) \,, \label{eq:d6b} \\
&&\psi_w \simeq  h(z)+{\cal O}(\rho^2) \,, \label{eq:d7b}
\end{eqnarray}
with $h(z)$ being some function of $ z$. We cannot determine how the other magnetic potential, $\psi_w$, behaves near the axises and hence do not require $h(z)_{[0]}=h(z)_{[1]}$ for the two solutions.

\medskip
We further consider the behavior of the magnetic potential $\mu$ defined 
by eq.~(\ref{eq:magnetic}). Since the norm of the rod vector $v$ vanishes 
over the outer axises, the first term in the right-hand side of 
eq.~(\ref{eq:magnetic}) vanishes there. 
Then, it follows from eq.~(\ref{eq:d6b}) that 
the derivative of the magnetic potential, $\mu$, is given by 
\begin{eqnarray}
d\mu=\mp\frac{c_\phi}{\sqrt{3}} dh(z).
\end{eqnarray} 
Integrating this, we obtain
\begin{eqnarray}
\mu=\mp\frac{c_\phi}{\sqrt{3}}  h(z)+c_1 \,, 
\end{eqnarray}
where $c_1$ is an integration constant. 
Here, note from eqs.~(\ref{eq:mu}) that $\mu=q/\sqrt{3}$ at $z=-\infty,\ \rho=0$ and  $\mu=-q/\sqrt{3}$ at $z=\infty,\ \rho=0$. 
Therefore, the constant $c_1$ is determined,
\begin{eqnarray}
c_1=\mp\frac{q}{\sqrt{3}}.
\end{eqnarray}
Thus, we can see that the magnetic potential, $\mu$, must 
behave as
\begin{eqnarray}
\mu\simeq \mp\frac{c_\phi}{\sqrt{3}}  h(z)\mp\frac{q}{\sqrt{3}}+{\cal O}({\rho^2}),\label{eq:outmu1}
\end{eqnarray}
near the outer axises.  

\medskip
Finally, let us consider the behaviors of the twist potentials $\omega_a$ near the outer axises.
From eqs.~(\ref{eq:d6b}) and (\ref{eq:d7b}), the derivatives of the twist 
potentials on the outer axises are give by 
\begin{eqnarray}
d\omega_a=\mp\frac{2c_\phi}{\sqrt{3}}  \psi_a dh(z) \,. 
\end{eqnarray}
Then, it follows that $\omega_a$ can be written 
\begin{eqnarray}
\omega_\phi=-\frac{2c_\phi^2}{3}h(z)+\frac{Nc_\phi}{\sqrt{3}} h(z)^2+c_2,\quad \omega_w=\mp\frac{c_\phi}{\sqrt{3}}  h(z)^2+c_3,
\end{eqnarray}
with $c_2$ and $c_3$ constants.
From eqs.(\ref{eq:ophi}) and (\ref{eq:opsi}), we easily find that 
\begin{eqnarray}
\omega_\phi=\pm 4 J+\frac{1}{2}(c_\phi q-N Q),\quad \omega_w=\mp Q \,.
\end{eqnarray}
at $\rho=0,\ z=\pm \infty$.
These boundary conditions at infinity for the twist potentials and $h(z=\pm\infty)=0$ determine the integration constants, $c_2$ and $c_3$, as 
\begin{eqnarray}
c_2=\pm 4J+\frac{1}{2}(c_\phi q -NQ),\quad c_3=\mp Q  \,.
\end{eqnarray}
Therefore, the twist potentials behave as 
\begin{eqnarray}
 \omega_\phi&=&-\frac{2c_\phi^2}{3}h(z)+ \frac{Nc_\phi}{\sqrt{3}} h(z)^2\pm 4J+\frac{1}{2}(c_\phi q -NQ),\label{eq:oomega1}\\
\omega_w&=&\mp\frac{c_\phi}{\sqrt{3}} h(z)^2\mp Q
\,.\label{eq:oomega2}
\end{eqnarray}
near the outer axises.

\medskip
Therefore, from eqs.(\ref{eq:olambda11})-(\ref{eq:olambda22}), (\ref{eq:d6b}), (\ref{eq:d7b}), (\ref{eq:outmu1}), (\ref{eq:oomega1}) and (\ref{eq:oomega2}), we can show that for $\rho \to 0$, $\rho\ \partial_z{\rm tr}\Psi$ behaves as
\begin{eqnarray}
\rho\ \partial_z{\rm tr}\ \Psi\simeq O(\rho) \,.  
\end{eqnarray}

Thus, we find from (i)--(iv) that the boundary integral, 
eq.~(\ref{eq:integral}), vanishes on each rod and the infinity. 
The deviation matrix, $\Psi$, is constant and has the asymptotic behavior, 
$\Psi\to 0$. Therefore, $\Psi$ vanishes over $\Sigma$, and the two 
configurations, $M_{[0]}$ and $M_{[1]}$, with the same values of constants, $(m,J,Q,N,q,c_\phi)$, must coincide with each other. 
This completes our proof for the uniqueness theorem for black holes.

\section{Boundary value problems for black rings}\label{sec:br}

In this section we would like to consider the boundary value problem for asymptotically Kaluza-Klein black rings.
In the Weyl-Papapetrou coordinate system, the boundaries for a black ring 
with the $S^1\times S^2$ horizon topology can be given as follows (See FIG.\ref{fig:rod}.(c) about the rod diagram):  
\begin{enumerate}
\item[(i)]
the outer axis: 
$\partial \Sigma_+=\{(\rho,z)|\rho=0,ck^2<z<\infty \}$ with the rod vector 
$v=(0,1,N)$ \,, 
\item[(ii)]
the inner axis 
$\partial \Sigma_{in}=\{(\rho,z)|\rho=0,k^2<z<ck^2 \}$ with the rod vector 
$v=(0,1,-N)$ \,,
\item[(iii)] the horizon: 
$\partial \Sigma_{\cal H}=\{(\rho,z)|\ \rho=0,-k^2<z<k^2\}$ \,, 
\item[(iv)] 
the outer axis
$\partial \Sigma_-=\{(\rho,z)|\rho=0,-\infty<z<-k^2\}$ 
with the rod vector $v=(0,1,-N)$ \,, 
\item[(v)] the infinity:  
$\partial \Sigma_\infty 
= \{(\rho,z)|\sqrt{\rho^2+z^2}\to 
\infty\ {\rm with}\ z/\sqrt{\rho^2+z^2}\ {\rm finite}  \}$ \,, \label{eq:idd}
\end{enumerate} 
where constants $c$ and $k$ satisfy $c>1$ and $0<k^2$. 

\medskip 
Therefore, the boundary integral in the left-hand side of the Mazur 
identity, eq.~(\ref{eq:mazurid}), is decomposed into the integrals over the four
rods (i)--(iv), and the integral at infinity (v), as 
\begin{eqnarray}
\int_{\partial \Sigma}\rho\partial_p{\rm tr}\Psi dS^p 
&=& \int_{-\infty}^{-k^2}\rho\frac{\partial {\rm tr}\Psi }{\partial z}dz 
   +\int_{-k^2}^{k^2}\rho\frac{\partial {\rm tr}\Psi }{\partial z}dz 
+\int_{k^2}^{ck^2}\rho\frac{\partial {\rm tr}\Psi }{\partial z}dz 
\nonumber\\
 && +\int_{ck^2}^{\infty}\rho\frac{\partial {\rm tr}\Psi }{\partial z}dz 
    +\int_{\partial\Sigma_\infty}\rho\partial_p{\rm tr}\Psi dS^p \,.   
\label{eq:descom}
\label{eq:integral} 
\end{eqnarray}

\medskip 
Note that the only difference between black holes and black rings appears at the third term in the right-side of eq.(\ref{eq:descom}),
which corresponds to the integral over the inner axis inside the black ring. As will be seen below, because of
the existence of this third integral, a dipole charge comes to appear in our boundary conditions.
For the boundaries ${\rm (i),(iii),(iv)}$ and ${\rm (v)}$, 
the boundary conditions of the scalar fields, $\Phi^A$, are exactly the 
same as those of black holes. Therefore, we consider only ${\rm (ii)}$.

\medskip
Noting that the rod vector is $v=(1,-N)$ for the inner axis,
we find that the regularity requires that the potentials, $\lambda_{ab}$, near the inner axis must behaves as
\begin{eqnarray}
&&\lambda_{\phi\phi}\simeq N^2k(z)+{\cal O}(\rho^2), \label{eq:ilambdaphiphi}\\
&&\lambda_{\phi w}\simeq  Nk(z)+{\cal O}(\rho^2), \label{eq:ilambdaphiw}\\
&&\lambda_{ww}\simeq k(z)+{\cal O}(\rho^2)\label{eq:ilambdaww}, 
\end{eqnarray}
where $k(z)$ is some function of $z$. 
The electric potentials satisfy 
\begin{eqnarray}
0=-i_vF=\sqrt{3}(d\psi_\phi-Nd\psi_w) \,. 
\end{eqnarray}
 Hence, integrating this, we obtain
\begin{eqnarray}
\psi_\phi-N\psi_w= c_{in} \,, 
\label{eq:cin}
\end{eqnarray}
where $c_{in}$ is an integration constant. 
Recall that the dipole charge, $q_m$, of a black ring is defined by
\begin{eqnarray}
q_m=\frac{1}{2\pi}\int_{S^2}F=\sqrt{3}\left[\psi_{\phi_-}(\rho=0,z=k^2)-\psi_{\phi_-}(\rho=0,z=-k^2)\right]=\sqrt{3}c_{in}+c_\phi.
\end{eqnarray}
Therefore, we see that the constant $c_{in}$ is related to the dipole charge $q$ by
\begin{eqnarray}
c_{in}=\frac{q_m-c_\phi}{\sqrt{3}}.
\end{eqnarray}
From eq. (\ref{eq:cin}) and the requirement of regularity, we can set the electric potentials, $\psi_a$,  to behave as
\begin{eqnarray}
&&\psi_\phi\simeq c_{in}+Nh(z)+{\cal O}(\rho^2), \label{eq:ipsiphi}\\
&&\psi_w\simeq h(z)+{\cal O}(\rho^2)\label{eq:ipsipsi}
\end{eqnarray}
in terms of some function $h(z)$ near the inner axis. Also note that in the boundary value analysis, we do not assume $h(z)_{[0]}=h(z)_{[1]}$.

\medskip
Next, let us see how the magnetric potential, $\mu$, behaves near the inner axis.
From eqs. (\ref{eq:magnetic}), (\ref{eq:ipsiphi}) and (\ref{eq:ipsipsi}), the derivative of the magnetic potential on the inner axis is written as
\begin{eqnarray}
d\mu=-c_{in} dh(z).
\end{eqnarray} 
Integrating this on the inner axis, we obtain
\begin{eqnarray}
\mu=-c_{in} h(z)+\tilde c_{1} \,, 
\end{eqnarray}
where $\tilde c_1$ is an integration constant. 
On the other hand, from eq. (\ref{eq:outmu1}), we note that just at the joint point $(\rho,z)=(0,ck^2)$ where the outer axis $\partial\Sigma_+$ and the inner axis $\partial\Sigma_{in}$ meet with each other, 
the magnetic potential takes the value of 
\begin{eqnarray}
\mu=-\frac{c_\phi}{\sqrt{3}}h(ck^2)-\frac{q}{\sqrt{3}}.
\end{eqnarray}
Hence, the continuity of the potential, $\mu$, at the point determines the value of the integration constant $\tilde c_1$, 
\begin{eqnarray}
\tilde c_1=\left(c_{in}-\frac{c_\phi}{\sqrt{3}}\right)h(ck^2)-\frac{q}{\sqrt{3}}.
\end{eqnarray}
Note from eqs. (\ref{eq:c0}) and (\ref{eq:cin}) and the continuity of the electric potentials that the equations, 
\begin{eqnarray}
&&\psi_\phi(\rho=0,z=ck^2)+N\psi_w(\rho=0,z=ck^2)=c_{in},\\
&&\psi_\phi(\rho=0,z=ck^2)-N\psi_w(\rho=0,z=ck^2)=\frac{c_\phi}{\sqrt{3}},
\end{eqnarray}
should hold at the point.
Solving these,  the value of $h(z)$ at $z=ck^2$ can be determined as
\begin{eqnarray}
h(ck^2)=\psi_w(\rho=0,z=ck^2)=-\frac{1}{2N}\left(c_{in}-\frac{c_\phi}{\sqrt{3}}\right)
\end{eqnarray}
in terms of the constants $N,c_\phi$ and $c_{in}$, {\it i.e.}, $N,c_\phi$ and $q_m$.
Therefore, the magnetic potential $\mu$ behaves as
\begin{eqnarray}
\mu\simeq -c_{in}h(z)-\frac{1}{2N}\left(c_{in}-\frac{c_\phi}{\sqrt{3}}\right)^2-\frac{q}{\sqrt{3}}+{\cal O}(\rho^2) \label{eq:imu}
\end{eqnarray}
near the inner axis.

\medskip
By the similar computations, we can see that the twist potentials behaves as
\begin{eqnarray}
&&\omega_\phi\simeq -2c_{in}^2h(z)+Nc_{in}h(z)^2+ \left[4J+\frac{1}{2}(c_\phi q-NQ)\right]+\tilde c_2+{\cal O}(\rho^2),\label{eq:iomegaphi}\\
&&\omega_w\simeq c_{in}h(z)^2- Q+{\cal O}(\rho^2)+\tilde c_3,\label{eq:iomegaw}
\end{eqnarray}
where the constants $\tilde c_2$ and $\tilde c_3$ are given by
\begin{eqnarray}
\tilde c_2&=&-\frac{1}{4N}\left(c_{in}-\frac{c_\phi}{\sqrt{3}}\right)^2\left(5c_{in}+\sqrt{3}c_\phi\right),\\
\tilde c_3&=&-\frac{1}{4N^2}\left(c_{in}-\frac{c_\phi}{\sqrt{3}}\right)^2\left(c_{in}+\frac{c_\phi}{\sqrt{3}}\right)
\end{eqnarray}

\medskip
Therefore, by using eqs. (\ref{eq:ilambdaphiphi})-(\ref{eq:ilambdaww}), (\ref{eq:ipsiphi}), (\ref{eq:ipsipsi}), (\ref{eq:imu}), (\ref{eq:iomegaphi}), (\ref{eq:iomegaw}), we can show that for $\rho \to 0$, $\rho\ {\rm tr}\Psi$ behaves as
\begin{eqnarray}
\rho\ \partial_z{\rm tr}\ \Psi\simeq O(\rho) \,.  
\end{eqnarray}

\medskip
Thus, we find that the boundary integral, 
eq.~(\ref{eq:integral}), vanishes on each rod and the infinity. 
The deviation matrix, $\Psi$, is constant and has the asymptotic behavior, 
$\Psi\to 0$. Therefore, $\Psi$ vanishes over $\Sigma$, and the two 
configurations, $M_{[0]}$ and $M_{[1]}$, with the same parameters $(m,J,Q,N,q,c_\phi,q_m)$ coincide with each other. 
This completes our proof for the uniqueness theorem for black rings.

\section{Boundary value problems for black lenses} \label{sec:bl}

Finally, let us consider the boundary value analysis for black lenses. 
In terms of the Weyl-Papapetrou coordinate system 
and the rod-structure \cite{Harmark}, 
the boundary $\partial \Sigma$ of the base space 
$\Sigma=\{(\rho,z)|\ \rho>0,\ -\infty<z<\infty \}$ is described as 
a set of three rods and the infinity: Namely, 
\begin{enumerate}
\item[(i)]
the outer axis: 
$\partial \Sigma_+=\{(\rho,z)|\rho=0,k^2<z<\infty \}$ with the rod vector 
$v=(0,1,N)$ \,, 
\item[(ii)] the horizon: 
$\partial \Sigma_{\cal H}=\{(\rho,z)|\ \rho=0,-k^2<z<k^2\}$ \,, 
\item[(iii)] 
the outer axis: 
$\partial \Sigma_-=\{(\rho,z)|\rho=0,-\infty<z<-k^2\}$ 
with the rod vector $v=(0,1-N)$ \,, 
\item[(iv)] the infinity:  
$\partial \Sigma_\infty 
= \{(\rho,z)|\sqrt{\rho^2+z^2}\to 
\infty\ {\rm with}\ z/\sqrt{\rho^2+z^2}\ {\rm finite}  \}$ \,. 
\end{enumerate} 
The above rod structure is similar to that of black holes but now the relation between the nut charge $N$ and the size of the 5-th dimension $L$ 
is given by $N=(L/2)n$. As mentioned in sec.~\ref{sec:infinity}, the spatial infinity is topologically a lens space $L(n;1)$ and hence from the absence of nuts 
 in the black hole exterior region, we can see that the topology of the horizon spatial cross section is $L(n;1)$. 
 It is clear that how to prove the uniqueness for the black lenses is entirely the same as the black hole case. 
Accordingly, we can conclude that
 the two configurations, $M_{[0]}$ and $M_{[1]}$, with the same parameters $(m,J,Q,N,q,c_\phi)$ coincide with each other.

\section{Summary and discussions}\label{sec:summary}

We have shown the uniqueness theorem which states that 
in five-dimensional minimal supergravity, stationary charged rotating black hole, or black lens  is 
uniquely specified by its asymptotic conserved charges and magnetic flux if (1) it admits 
two independent axial Killing symmetries, (2) the event 
horizon cross-section is connected and non-degenerate 
(3) there are not any nut and any bolt in the domain of outer communication. 
Furthermore, we have also shown that under the assumptions (1) and (2), stationary charged rotating black ring with an event
horizon of the cross-section topology $S^1 \times S^2$ is 
uniquely specified by the dipole charge and rod structure in addition its asymptotic conserved charges and magnetic flux.
Our theorem generalizes the uniqueness theorem for Kaluza-Klein black 
holes in five-dimensional vacuum Einstein gravity \cite{MI}, or in five-dimensional Einstein-Maxwell theory~\cite{Y10} to 
the case of five-dimensional minimal supergravity. 

\medskip
Finally, we comment on the assumption (3) in our proof.
This assumption (3) restricts the topologies of the black hole exterior regions to the simplest cases.  
When there exists a nut, or a bolt ---joint points of two spacelike rods--- outside the horizon, the rod structure can have the isolated and finite spacelike rod
 which cannot be connected with infinity. We here call it {\it inner axis}. 
As seen in the proof of black rings, the integration constant $c'$ which is defined by $\psi_\phi+N'\psi_w=c'$ 
 appears in the boundary condition on the inner axis.
We have not been able to relate the integration constant to any of the other charges, except for the vacuum case $(q = c_\phi = 0)$. We also
see that a similar problem just mentioned above occur when we consider uniqueness theorems for multi-rings,
black Saturn, or more complicated black objects. This issue deserves to further study.

\section*{Acknowledgments} 

We would like to thank A. Ishibashi for valuable discussions and 
comments. S.T. is supported by the JSPS under Contract No. 20-10616.

\section*{Appendix: 
Coset matrix and the Mazur identity}\label{sec:mazur} 
Here, to be self-contained, we provide the coset matrix representation 
and the Mazur identity for our non-linear sigma model. 

\medskip 
Following ~\cite{BCCGSW}, we introduce the $G_{2(2)}/SO(4)$ coset matrix, 
$M$, defined by 
\begin{eqnarray}
M= \left(
  \begin{array}{ccc}
  \hat A&\hat B&\sqrt{2}\hat U\\
  \hat B^T&\hat C&\sqrt{2}\hat V\\
  \sqrt{2}\hat U^T&\sqrt{2}\hat V^T&\hat S\\
  \end{array}
 \right) \,,
\end{eqnarray}
where $\hat A$ and $\hat C$ are symmetric $3\times 3$ matrices, $\hat B$ is a $3\times 3$ 
matrix, $\hat U$ and $\hat V$ are 3-component column matrices, and $\hat S$ is a scalar, 
defined, respectively, by  
\begin{eqnarray}
&&\hat A=\left(
  \begin{array}{ccc}
  [(1-y)\lambda+(2+x)\psi \psi^T-\tau^{-1}\tilde\omega\tilde\omega^T+\mu(\psi \psi^T\lambda^{-1}\hat J-\hat J\lambda^{-1}\psi\psi^T)]&\tau^{-1}\tilde\omega\\
 \tau^{-1}\tilde\omega^T& -\tau^{-1}
  \end{array}
 \right) \,,\nonumber\\
&&\hat B=\left(
  \begin{array}{ccc}
  (\psi\psi^T-\mu \hat J)\lambda^{-1}-\tau^{-1}\tilde\omega \psi^T \hat J&[(-(1+y)\lambda \hat J-(2+x)\mu+\psi^T\lambda^{-1}\tilde\omega)\psi+(z-\mu \hat J\lambda^{-1}\tilde)\omega] \\
  \tau^{-1}\psi^T \hat J&-z\\
  \end{array}
 \right) \,, \nonumber\\
&&\hat C=\left(
  \begin{array}{ccc}
 (1+x)\lambda^{-1}-\lambda^{-1}\psi\psi^T\lambda^{-1}&\lambda^{-1}\tilde\omega-\hat J(z-\mu \hat J \lambda^{-1})\psi\\
 \tilde\omega^T\lambda^{-1}+\psi^T(z+\mu \lambda^{-1}\hat J)\hat J&[\tilde\omega^T\lambda^{-1}\tilde\omega-2\mu\psi^T\lambda^{-1}\tilde\omega-\tau(1+x-2y-xy+z^2)]\\
  \end{array}
 \right) \,,\nonumber\\
&&\hat U=
\left(
  \begin{array}{cc}
  (1+x-\mu \hat J\lambda^{-1})\psi-\mu \tau^{-1}\tilde\omega\\
  \mu\tau^{-1}\\
  \end{array}
 \right) \,,\nonumber\\
&&\hat V=\left(
  \begin{array}{ccc}
  (\lambda^{-1}+\mu\tau^{-1}\hat J)\psi\\
  \psi^T\lambda^{-1}\tilde\omega-\mu(1+x-z)\nonumber\\
  \end{array}
 \right) \,,\nonumber\\
&&\hat S=1+2(x-y) \,, \nonumber
\end{eqnarray}
with 
\begin{eqnarray}
&&\tilde \omega=\omega-\mu\psi \,,
\end{eqnarray}
\begin{eqnarray}
&&x=\psi^T\lambda^{-1}\psi,\quad y 
   =\tau^{-1}\mu^2,\quad z=y-\tau^{-1}\psi^T\hat J\tilde\omega \,,
\end{eqnarray}
and the $2\times 2$ matrix, 
\begin{eqnarray}
\hat J= \left(
  \begin{array}{ccccccc}
   0&1\\
   -1&0\\
  \end{array}
 \right) \,.  
\end{eqnarray}
We note that this $7\times7$ matrix $M$ is symmetric, $M^T=M$, 
and unimodular, $\det(M)= 1$. Since we choose the Killing vector fields 
$\xi_\phi$ and $\xi_w$ to be spacelike, all the eigenvalues of $M$ are 
real and positive. Therefore, there exists an $G_{2(2)}$ matrix $\hat g$ such that 
\begin{eqnarray}
  M = \hat g \hat g^T \,.
\end{eqnarray} 
We define a current matrix as
\begin{eqnarray}
 J_i = M^{-1} \partial_i M \,,
\end{eqnarray}
which is conserved if the scalar fields are the solutions of the equation of motion derived by the action (\ref{action}). Then, the action (\ref{action}) can be written in terms of $J$ and $M$ as follows
\begin{eqnarray}
S&=&\frac{1}{4}\int d\rho dz \rho {\rm tr}(J_iJ^i) \nonumber \\
 &=&\frac{1}{4}\int d\rho dz 
    \rho {\rm tr}(M^{-1}\partial_iMM^{-1}\partial^iM)\,. \label{eq:action2}
\end{eqnarray}
Thus, the matrix $M$ completely specify the solutions to our system.   

\medskip 
Let us now consider two sets of field configurations, 
$M_{[0]}$ and $M_{[1]}$, that satisfy the equations of motion derived 
from the action, eq.~(\ref{eq:action2}). 
We denote the difference between the value of the functional obtained from 
the field configuration $M_{[1]}$ and the value obtained from $M_{[0]}$ 
as a bull's eye $\stackrel{\odot}{}$, e.g., 
\begin{eqnarray}
\stackrel{\odot}J{}^i=J^i_{[1]}-J^i_{[0]} \,,
\end{eqnarray}
where the subscripts ${}_{[0]}$ and ${}_{[1]}$ denote, respectively, 
the quantities associated with the field configurations $M_{[0]}$ 
and $M_{[1]}$. The deviation matrix, $\Psi$, is then defined by
\begin{eqnarray}
\Psi=\stackrel{\odot}MM^{-1}_{[0]}=M_{[1]}M^{-1}_{[0]}-{\bf 1} \,,
\end{eqnarray}
where ${\bf 1}$ is the unit matrix. Taking the derivative of this, 
we have the relation between the derivative of the deviation matrix 
and $\stackrel{\odot}J{}^i$, 
\begin{eqnarray}
D^i\Psi=M_{[1]}\stackrel{\odot}J{}^iM_{[0]}^{-1} \,, 
\label{eq:deriv}
\end{eqnarray}
where $D_i$ is a covariant derivative associated with the abstract 
three-metric $\gamma$. 
Taking, further, the divergence of the above formula and also the trace of 
the matrix elements, we have the following divergence identity 
\begin{eqnarray}
D_i D^i {\rm tr} \Psi 
 = {\rm tr} \left( \stackrel{\odot}J{}^{Ti} M_{[1]}\stackrel{\odot}J{}^iM_{[0]}^{-1} \right) \,,   
\label{id:global-divergence}
\end{eqnarray}
where we have also used the conservation equation $D_i J{}^i =0$. 
Then, integrating this divergence identity over the region 
$\Sigma=\{(\rho,z)|\rho\ge 0,\ -\infty<z<\infty \}$, we obtain 
the Mazur identity, 
\begin{eqnarray}
\int_{\partial \Sigma}\rho \partial_p {\rm tr} \Psi dS^p 
= \int_{\Sigma}\rho \hat h_{pq}{\rm tr} 
 ({\cal M}^{Tp} \: {\cal M}^q)d\rho dz \,, 
\label{eq:id} 
\end{eqnarray}
where $\hat h_{pq}$ is the two-dimensional flat metric 
\begin{eqnarray}
\hat h=d\rho^2+dz^2 \,, 
\end{eqnarray}
and the matrix ${\cal M}$ is defined by 
\begin{eqnarray}
{\cal M}^p=\hat g_{[0]}^{-1}\: \stackrel{\odot}J{}^{Tp} \: \hat g_{[1]} \,. 
\end{eqnarray}

\end{document}